\documentclass[12pt, a4paper] {article}

\usepackage{amsfonts}
\usepackage{verbatim}
\usepackage{amsmath}
\usepackage{color}

\begin{document}

\begin{center}

{\sf {\Large Curci-Ferrari-type condition in Hamiltonian formalism:\\ A free spinning relativistic particle }}

\vskip 3.0 cm

{\sf{ A. Shukla$^{(a)}$, T. Bhanja$^{(a)}$, R. P. Malik$^{(a,b)}$}}\\
{\it $^{(a)}$Department of Physics, Centre of Advanced Studies,}\\
{\it Banaras Hindu University, Varanasi - 221 005, (U. P.), India}\\

\vskip 0.1cm

{\bf and}\\

\vskip 0.1cm

{\it $^{(b)}$DST Centre for Interdisciplinary Mathematical Sciences,}\\
{\it Faculty of Science, Banaras Hindu University, Varanasi - 221 005, India}\\

\end{center}

\vskip 1.5 cm

\noindent

\noindent
{\bf Abstract:} The Curci-Ferrari (CF)-type restriction emerges in the description of a free spinning relativistic 
particle within the framework of Becchi-Rouet-Stora-Tyutin (BRST) formalism when the off-shell nilpotent and 
absolutely anticommuting (anti-)BRST symmetry transformations for this system are derived from the application of horizontality 
condition (HC) and its supersymmetric generalization (SUSY-HC) within the framework of superfield formalism. We show that the above 
CF condition, which turns out to be the secondary constraint of our present theory, remains time-evolution invariant 
within the framework of Hamiltonian formalism. This time-evolution invariance (i) physically justifies 
the imposition  of the (anti-)BRST       
invariant CF-type condition on this system, and 
(ii) mathematically implies the linear independence of BRST and anti-BRST 
symmetries of our present theory.\\ 



\newpage

\section{Introduction}
The standard model of high energy physics is one of the most successful theories of modern times
which is based on the four (3 + 1)-dimensional (4D) non-Abelian 1-form ($A^{(1)} = dx^\mu A_\mu;\, \mu =0, 1, 2, 3 $) 
gauge theory. Such theories are covariantly quantized by using Becchi-Rouet-Stora-Tyutin (BRST)
formalism where the unitarity and ``quantum" gauge (i.e. BRST) 
invariance are respected {\it together} at any arbitrary order
of perturbative computations for a given physical process (that is allowed by the interacting gauge theory).

One of the decisive features of BRST approach to any arbitrary $p$-form ($p = 1, 2, 3...$) {\it non-supersymmetric}
gauge theories is the existence of Curci-Ferrari (CF) type restriction{\footnote{For the 
Abelian 1-form gauge theory, the CF-type restriction turns out to be {\it trivial}
(as it is the limiting case of CF condition present in its non-Abelian counterpart).}}
(see, e.g. [1]) which has been recently shown to
be deeply connected with the mathematical concept of gerbes (see, e.g. [2,3]). This condition ensures the absolute 
anticommutativity of BRST and anti-BRST symmetries thereby making the BRST and anti-BRST 
symmetries have completely independent identity and meaning.

The geometrical superfield formalism [4,5] provides a sound mathematical basis for the derivation
of CF-type restriction for any arbitrary $p$-form non-supersymmetric gauge theory 
where the celebrated horizontality condition (HC) plays a very decisive role. In our recent publication [6],
we have applied the superfield formulation to a toy model of {\it supersymmetric} (SUSY) gauge theory 
(i.e. a free spinning relativistic particle) where we have generalized HC to its supersymmetric form (SUSY-HC).
The application of the latter leads to the emergence of a CF-type restriction [cf. equation (6) below]
and proper (anti-)BRST symmetries [cf. equations (2) and (3) below]. The Noether conserved charges,
corresponding to the latter symmetries, have been derived and some novel features in the physicality
criteria (with the conserved and nilpotent (anti-)BRST charges) have been pointed out in [7].

In the above context, it is important to stress that {\it neither} the Lagrangian 
{\it nor} the superfield formalism answers the question as to why the CF-type restriction should be 
imposed on the theory within the framework of BRST formalism. The central purpose of our present 
investigation is to answer the above question by
demonstrating the time-evolution invariance of CF-type restriction within the
framework of Hamiltonian formalism so that the imposition of the latter (on our physical system) could 
be physically justified. We accomplish this goal by taking the help of appropriate Lagrangian and corresponding
Hamiltonian for the SUSY system of spinning relativistic particle. To be precise, we redefine the specific auxiliary 
variables of the theory to achieve our central goal in one stroke [cf. (24), (28) and (30) below].

The motivating factors behind our present investigation are as follows. First, the Lagrangian
and superfield formulations [4,5] do not shed any light on the time-evolution invariance of the
CF-type condition. Thus, it is important for us to prove it within the framework of Hamiltonian
formulation. Second, for aesthetic reasons, it is essential to obtain the CF-type condition
from a single Lagrangian and/or Hamiltonian. We have accomplished this goal in our present 
endeavor. Finally, our present work is a modest step towards our main goal of applying the
superfield and BRST formulations to the SUSY $p$-form gauge theories of phenomenological importance.

Our present paper is organized as follows. In section two, we concisely recapitulate
the bare essentials of the proper (i.e. off-shell nilpotent and absolutely anticommuting)
(anti-)BRST symmetries, coupled Lagrangians and CF-type restriction. Our section three is 
devoted to the derivation of alternative sets of Lagrangians (and Hamiltonians) that are
useful in the description of spinning relativistic particle and which yield the CF-type
restriction in one step. Section four deals with the time-evolution invariance
of the CF-type restriction within the framework of Hamiltonian formalism. Finally, in section
five, we make some concluding remarks.

\section{Preliminaries: Off-shell nilpotent symmetries in the Lagrangian formulation} 

We begin with the one $(0+1)$-dimensional (1D) SUSY system of a massless spinning particle
 whose Lagrangian is (see, e.g. [8])
\begin{eqnarray}
 L_0 = p_\mu\; \dot x^\mu - \frac{e}{2}\, p^2 + \frac{i}{2}\, \psi_\mu\, \dot \psi^\mu + i \,\chi\, p_\mu\, \psi^\mu,
\end{eqnarray}
where the motion of the 1D SUSY particle is parameterized by the
monotonically increasing 
parameter $\tau$ and $\dot x^\mu = (dx ^\mu /d \tau)$ and $\dot \psi ^\mu = ({d} {\psi ^\mu}/ d\tau) $
are the generalized "velocities" corresponding to the $D$-dimensional target space position variable
 $x^\mu(\tau) \,( \mu = 0,1,2...D-1)$ and its supersymmetric partner $\psi^ \mu (\tau)$. The latter 
turns out to be the spin variable of this particle. 
The constraints $p^2 = 0$ and $p_\mu\psi^\mu =0$ of our present theory 
have been incorporated through the Lagrange multipliers  $e(\tau)$ and $\chi(\tau)$ 
which are bosonic and fermionic in nature, respectively.
These latter variables are analogues of the vierbein and Rarita-Schwinger fields of the 4D supergravity theory.
In addition, these variables transform exactly like the gauge potentials of the 4D supersymmetric gauge theories.
The fermionic variables ($\psi_\mu,\, \chi$) anticommute (i.e. $\chi^2 = 0,\, \psi^2_\mu = 0, \,
\psi_\mu \psi_\nu + \psi_\nu \psi_\mu = 0,\, \psi_\mu \chi +\chi \psi_\mu = 0,$ etc.) amongst themselves 
and commute (i.e. $\psi_\mu x_\nu - x_\nu \psi_\mu = 0,\,  \chi x_\mu - x_\mu \chi = 0,\,
e \chi - \chi e= 0,\, \psi_\mu e - e \psi_\mu = 0,$ etc.) with the bosonic variables ($x_\mu,\, e,\, p_\mu,$ etc.) of the theory.
The basic fermionic ($\psi_\mu,\,\chi $) and bosonic ($x_\mu,\, e $) degrees of freedom match which
is the indication of presence of supersymmetry in the theory.

The well-known gauge and supergauge symmetry transformations [8] 
of the above Lagrangian can be generalized  to the proper (i.e. off-shell nilpotent and absolutely anticommuting)
(anti-)BRST symmetry transformations $(s_{(a)b})$ at the quantum level of the theory, as (see, e.g. [6] for details)
\begin{eqnarray} 
&& s_{ab}\; x_\mu = {\bar c}\; p_\mu + \bar \beta \;\psi_\mu,\qquad s_{ab}\; e = \dot {\bar c} + 2 \;\bar \beta\; \chi,  
\;\qquad s_{ab} \;\psi_\mu = i \;\bar \beta\; p_\mu,\nonumber\\
&& s_{ab}\;\bar c = - i \;{\bar \beta}^2,\quad s_{ab}\; c = i\; \bar b, \quad s_{ab}\; \bar \beta = 0, 
\quad s_{ab} \; \beta = - i\; \gamma,\quad s_{ab}\;p_\mu = 0, \nonumber\\
&& s_{ab} \;\gamma = 0,\quad s_{ab}\; \bar b = 0, \quad s_{ab}\;\chi = i\; \dot {\bar \beta}, 
\quad s_{ab} \; b =  2\; i\; \bar \beta\; \gamma,
\end{eqnarray}
\begin{eqnarray}
&&s_b\; x_\mu = c\;p_\mu + \beta \;\psi_\mu,\qquad s_b\; e = \dot c + 2\;\beta\; \chi,  
\qquad s_b\; \psi_\mu = i\;\beta\; p_\mu,\nonumber\\
&& s_b\;c = - i\; \beta^2, \;\quad s_b \;{\bar c} = i\; b, \;\quad s_b \;\beta = 0, 
\;\quad s_b \;\bar \beta = i \;\gamma, \;\quad s_b\; p_\mu = 0,\nonumber\\
&& s_b \;\gamma = 0, \qquad s_b \;b = 0, \qquad s_b \;\chi = i\; \dot \beta, 
\qquad s_b\; \bar b = - 2\; i\; \beta\; \gamma,
\end{eqnarray}
which are the symmetry transformations for the following coupled (but  equivalent) Lagrangians
of our present physical system [6]
\begin{eqnarray}
L_b &=& L_0 + b \;\dot e + b(b+ 2\;\beta \bar \beta) - 
i \;\dot {\bar c}\; (\dot c + 2\; \beta\; \chi) 
+ 2 \; i\; \bar \beta\;  \dot c\; \chi\nonumber\\
&-& 2\; e\; (\gamma\; \chi + \bar \beta\; \dot \beta) + 2\; \beta \; \gamma\; \bar c 
+ \bar \beta ^2\; \beta^2 + 2\; \bar\beta\;\ c \; \gamma,
\end{eqnarray}
\begin{eqnarray}
L_ {\bar b} &=&L_0 - \bar b \;\dot e + \bar b(\bar b+ 2\;\bar \beta\;  \beta) - 
i \;\dot {\bar c}\; (\dot c + 2\; \beta\; \chi) 
+ 2 \; i\; \bar \beta\;  \dot c\; \chi\nonumber\\
&-& 2\; e\; (\gamma\; \chi - \beta\; \dot {\bar\beta} )
+ 2\; \beta \; \gamma\; \bar c + \bar \beta ^2\; \beta^2 + 2\; \bar\beta\;\ c \; \gamma,
\end{eqnarray}
where $(b,\; \bar b)$ are the Nakanishi-Lautrup-type auxiliary variables, $(\bar c)c$ are
the fermionic (anti-)ghost variables (i.e. $ c^2	=	\bar c^2 =0, \;c\,\bar c + \bar c\, c = 0 )$ and 
($\bar \beta)\beta$ are the bosonic (anti-)ghost variables. The fermionic (i.e. $\gamma \chi + \chi \gamma = 0,\,
\gamma \psi_\mu + \psi_\mu \gamma = 0 $) variable $\gamma$ is also included in the theory for its complete
and consistence description within the framework of BRST formalism.    
The above (anti-)ghost variables are required for the validity of unitarity in the theory.
One of the key observations, from the coupled Lagrangians (4) and (5), is the following
equations of motion:
\begin{eqnarray}
b = - \frac {\dot {e}}{2} - \beta\, \bar \beta, \qquad \bar b = \frac {\dot {e}}{2} 
- \beta\, \bar\beta \,\quad\Longrightarrow \quad b + \bar b + 2\,\beta\, \bar \beta = 0.
\end{eqnarray}
The above final expression $ b + \bar b + 2\,\beta\,\bar\beta = 0$ is nothing but the Curci-Ferrari (CF)-type
restriction that is responsible for the absolute anticommutativity of the (anti-)BRST symmetry transformations.
For instance, the basic anticommutators $\{ s_b, s_{ab}\}\,e =0$ and $\{ s_b,s_{ab} \}\,x_ \mu = 0$  
are true if and only if  $ b + \bar b + 2\beta \bar \beta = 0 $. The existence of the CF-type restriction,
in the context of spinning relativistic particle, is a completely {\it novel} observation which has emerged out
from the description of our present SUSY system within the framework of superfield formalism (see, e.g. [6] for details).

We also note that the (anti-)BRST invariant (i.e. $s_{(a)b}\, [ b + \bar b + 2\,\beta\,\bar\beta = 0]$)
CF-type restriction has been obtained in two steps because first  we 
have derived the expressions for $ b $ and $\bar b$  from Lagrangians (4)  and (5) and, then, we have added 
them together to obtain (6). Further, it is {\it not} clear as to why one should impose the 
CF-type condition for the absolute anticommutativity property (within the framework of Lagrangian formalism). 
Thus, in the forthcoming sections, we shall discuss the time-evolution invariance of CF-type restriction 
within the framework of Hamiltonian formulation so that we could physically justify its imposition 
(during the complete time-evolution).

\section{Lagrangians and corresponding canonical Hamiltonians: alternative forms}

We discuss, in this section, the Lagrangian and Hamiltonian formulations of our present theory in an explicit fashion.
In this context, first of all, we re-express the coupled Lagrangians [cf. (4),(5)] in the following forms
\begin{eqnarray}
L^{(1)}_b &=& L_0 + b \,\dot e + b(b+ 2\;\beta \bar \beta) - 
i \,\dot {\bar c}\, \dot c + \bar \beta ^2\, \beta^2 - 2\,i\, \dot{\bar\beta}\,c\, \chi
+ 2\,i\,(\beta\,\bar c - \bar \beta\,c)\,\dot \chi\nonumber\\
&+& 2\,i\,(\bar c \,\chi + i\,e\,\bar\beta)\,\dot \beta + 2\,( e\,\chi - \beta\,\bar c + \bar\beta\,c)\,\gamma,
\end{eqnarray}
\begin{eqnarray}
L^{(1)}_{\bar b} &=& L_0 -  \bar b \,\dot e + \bar b(\bar b+ 2\;\beta \bar \beta) - 
i \,\dot {\bar c}\, \dot c + \bar \beta ^2\, \beta^2 - 2\,i\, \dot{\bar\beta}\,c\, \chi
+ 2\,i\,(\beta\,\bar c - \bar \beta\,c)\,\dot \chi\nonumber\\
&-& 2\,i\,( c \,\chi + i\,e\,\beta)\,\dot {\bar\beta} + 2\,( e\,\chi - \beta\,\bar c + \bar\beta\,c)\,\gamma,
\end{eqnarray}
which differ from (4) and (5) by the total derivative, namely,
\begin{eqnarray}
L_{(b,\bar b)} = L^{(1)}_{(b,\bar b)} + \frac {d}{d\tau} \Big[2\,i\,\bar\beta \,c\,\chi 
- 2\,i\,\beta\, \bar c\, \chi \Big].
\end{eqnarray}
As a consequence, the dynamical equations of motion for the  theory remain intact (because the
action remains the same for physical variables which vanish off at infinity).
In our further discussions, the forms (7) and (8) would be preferred over (4) and (5) because the former 
coupled set supports the maximum number of variables having their non-vanishing conjugate momenta.

Let us first concentrate on the Lagrangian (7) which is {\it perfectly} BRST invariant [6].
The Euler-Lagrange equations of motion, from (7), are 
\begin{eqnarray}
&&{\dot x}_\mu - e \, p_\mu + i\, \chi\, \psi_\mu = 0,  \qquad
\dot \psi_\mu - \chi\, p_\mu = 0,\qquad \dot e + 2\,(b + \beta\, \bar\beta ) = 0, \nonumber\\
&& \dot b + \frac {p^2}{2} + 2\, \gamma\,\chi + 2\,{\dot \beta} \, \bar\beta = 0, \qquad 
p_\mu\, \psi^\mu + 2\, \dot {\bar c}\, \beta - 2\, \bar\beta\, \dot c - 2\,i\,e\, \gamma = 0, \nonumber\\
&&\ddot {c} + 2\, \dot {\beta}\,\chi + 2\, \beta\,\dot {\chi} + 2\,i\,\beta\, \gamma = 0, \qquad 
\ddot {\bar c} + 2\, \dot {\bar \beta}\,\chi + 2\, \bar\beta\,\dot {\chi} + 2\,i\,\bar\beta\, \gamma = 0, \nonumber\\
&&{\dot p}_\mu = 0, \qquad e\,\chi - \beta\, \bar c + \, \bar\beta\,c = 0, \qquad b\,\beta + i\, \dot{c}\,\chi - e \,\dot{\beta} 
+ \beta\, {\bar\beta}^2 + c\,\gamma = 0, \nonumber\\ 
&&\dot {e}\,\bar\beta + e\, \dot {\bar\beta} + b\,\bar\beta - i\, \dot{\bar c}\,\chi + \gamma\, \bar c
+ \beta\,\bar\beta^2 = 0,
\end{eqnarray}
where we have used the convention of the ``left-derivative" in the operation of derivatives w.r.t. the 
fermionic variables.
Exactly the same kind of equations of motion emerge out from the Lagrangian (8) except the following 
\begin{eqnarray}
&&\dot e - 2\,(\bar b + \beta\, \bar\beta ) = 0, \qquad \dot {e}\,\beta + e\, \dot {\beta} 
- \bar b\,\beta - i\, \dot{c}\,\chi + \gamma\, c - \beta^2\,\bar\beta = 0 , \nonumber\\
&& \dot {\bar b} - \frac {p^2}{2} - 2\, \gamma\,\chi + 2\,\beta \,\dot {\bar\beta} = 0,
\qquad e\,\dot{\bar\beta} - i\, \dot{\bar c}\,\chi - \bar c\,\gamma + \bar b\, \bar\beta  = 0.
\end{eqnarray}
It is to be noted that the equations of motion, corresponding to $b$ and $\bar b$ in (10) and (11), lead to
the derivation of CF condition (6). In addition, the variation of the action with 
respect to the variable $e(\tau)$ also
yields the CF-type condition (6). At this juncture, it is crystal clear that the CF-type condition (6) is derived
in {\it two steps} by adding appropriate equations of motion from (10) and (11) which are true
for the Lagrangians (7) and (8).

Using the Legendre's transformation, one can calculate the Hamiltonian, corresponding to the Lagrangians (7) 
and (8), as given below
\begin{eqnarray}
H_{b} &=& i\, \Pi_{(\bar c)}\, \Pi_{(c)} + \frac{e}{2}\,p^2 - i\,\chi(p_\mu\, \psi^\mu) - b (b + 2\,\beta\,\bar\beta) 
-\bar\beta^2\, \beta^2 \nonumber\\ &-& 2\,e\,\chi\,\gamma + i\, \Pi_{(\chi)}\,\gamma,
\end{eqnarray}
\begin{eqnarray}
H_{\bar b} &=& i\, \Pi_{(\bar c)}\, \Pi_{(c)} + \frac{e}{2}\,p^2 - i\,\chi(p_\mu\, \psi^\mu) -\bar b (\bar b + 2\,\beta\,\bar\beta) 
- \bar\beta^2\, \beta^2 \nonumber\\&-& 2\,e\,\chi\, \gamma + i\, \Pi_{(\chi)}\,\gamma ,
\end{eqnarray}
where we have used $H_{(b,\bar b)} = \dot {\phi}_i\, \Pi^i_{(\phi)} - L_{(b, \bar b)}$ for the
generic variable $\phi_i$ (i.e. $\phi_i = x_\mu,\, \psi_\mu, \, e,\, \chi,\, \beta,\,\bar\beta, \, c,\, \bar c,\, \gamma$)
and corresponding momenta $\Pi^i_{(\phi)}$. In fact, the explicit forms of the canonically conjugate momenta  
($\Pi^i_{(\phi)}$) are 
\begin{eqnarray}
&&\Pi^{\mu}_{(x)} = p^{\mu}, \quad \Pi^{\mu}_{(\psi)} = -\frac{i}{2}\,\psi^{\mu},
\quad \Pi_{(e)} = b,\nonumber\\ && \Pi_{(\chi)} = -2\,i\,(\beta\,\bar c - \bar\beta\, c), \qquad
\Pi_{(\beta)} = 2\,i\,(\bar c\, \chi +i\, e\, \bar\beta), \nonumber\\ &&\Pi_{(\beta)} = -2\,i\,c\,\chi, 
\qquad \Pi_{(c)} = i\, \dot{\bar c}, \qquad \Pi_{(\bar c)} = -i\; \dot c, 
\end{eqnarray}
which have been derived from the Lagrangian (7). The conjugate momenta, from the Lagrangian (8), are 
also the {\it same} except the following 
\begin{eqnarray}
\Pi_{(e)} = - \bar b, \qquad \;\Pi_{(\beta)} = 2\,\,i\, \bar c\, \chi, \qquad 
\Pi_{(\bar\beta)} = - 2\,i\, (c\,\chi + i\,e\, \beta ). 
\end{eqnarray}
It should be noted that $\Pi_{(\gamma)} \approx 0$ is the primary constraint on the theory 
and $\dot {\Pi}_{(\gamma)} \approx 0$ (calculated from $H_{(b, \bar b)}$) leads to $e\,\chi = \beta\, \bar c - \bar\beta\, c$
[cf. (10) as well]. In fact, one can add (i.e. $H_T = H_{(b, \bar b)} + v\,\Pi_{\gamma}$) the primary
constraints $\Pi_{\gamma}$ in the Hamiltonians $H_{(b, \bar b)}$ following Dirac's prescription.
However, the equation of motion from $H_T$ imply that the arbitrary coefficient $v = \dot{\gamma}$ (so that $H_T = H{(b, \bar b)} 
+ \dot {\gamma} \, \Pi_\gamma$). We have purposely ignored the presence of $\dot{\gamma}\, \Pi_{(\gamma)}$
in the above Hamiltonian mainly to avoid the ``velocity'' dependence.

We can use the following Heisenberg's equations of motion (with $\hbar = 1$) for the generic variable $\phi_i$ 
(with  generic entity $\phi_i$ corresponding to all the variables and corresponding momenta), namely;
\begin{eqnarray}
\dot {\phi_i} = \pm\,i\,\Bigl [\phi, H_{(b, \bar b)} \Bigr ], \quad \qquad\;\; \dot {\phi_i} = \frac{d\phi_i}{d\tau},
\end{eqnarray}
where ($\pm$) signs, in front of the commutator, are chosen depending on the nature of the generic variable
$\phi_i$ being (fermionic)bosonic in nature, to derive the dynamics of the theory. In fact, we obtain the 
following Heisenberg's equations of motion from the Hamiltonian $H_b$, namely;
\begin{eqnarray}
&& e\,\chi = \beta\,\bar c - \bar\beta\, c, \quad \dot \chi = - i\,\gamma, \quad
\dot {\Pi}_{(\chi)} = i\,p_\mu \, \psi^\mu + 2\,e\, \gamma,\quad \dot \beta = \dot{\bar\beta} = 0,\nonumber\\
&& \ddot c = \ddot{\bar c} = 0,\quad \Pi^\mu_{(x)} \equiv \dot p^\mu = 0, \quad \dot {x}_\mu = e\,p_\mu - i\, \chi\, \psi_\mu,
\quad \dot {\psi}_\mu = \chi\, p_\mu,\nonumber\\
&& \dot{\bar c} = \dot {\bar c}, \quad \dot {\Pi}_{(\beta)} = 2\, \beta\, \bar\beta^2 + 2\, b\, \bar\beta,
\quad  \dot e + 2\,b + 2\,\beta\,\bar\beta = 0, \quad \dot \gamma = \dot \gamma, \nonumber\\
&&\dot c =\dot c, \qquad \dot b + \frac{1}{2}  p^2 + 2 \gamma \chi = 0,
\quad \dot {\Pi}_{(\bar\beta)} = 2\,\bar\beta\,\beta^2 + 2\, b\, \beta,
\end{eqnarray}
where we have used the canonical brackets as illustrated below 
\begin{eqnarray}
&&\big[e, \Pi_e \big] = i, \qquad \{\chi, \Pi_{(\chi})\} = i, \qquad \{c, \Pi_{(c)}\} = i, \quad  \{\bar c, \Pi_{(\bar c)} \} = i, \nonumber\\
&&\big[x_{\mu} , p^{\nu}\big] = i\,\delta_{\mu}^{\nu}, \qquad
 \{ \psi_\mu, \psi^\nu\} = - \delta_{\mu}^\nu, \qquad [\beta, \Pi_{(\beta)}] = i,\nonumber\\ 
&&\big[\bar\beta, \Pi_{(\bar\beta)}\big] = i, \qquad \{ \gamma, \Pi_{\gamma}\} = i,
\end{eqnarray}
and the rest of the brackets have been taken to be zero. We note{\footnote{ We choose our target spacetime D-dimensional manifold to
be Minkowskian with flat metric $\eta_{\mu\nu} =$ diag (+1, -1,...) where
$\eta_{\mu\nu} \eta^{\nu\lambda} = \delta^{\lambda}_{\mu}$ and the dot product $P \cdot Q = \eta_{\mu\nu} P^\mu Q^\nu 
\equiv P^\mu\,Q_\mu$ for the non-null vectors $P_\mu$ and $Q_\mu$ in the D-dimensional target space.}}
that we have: $\Pi^\mu_{(\psi)} = -\frac{i}{2}\, \psi^\mu$
and $\{\psi_\mu, \psi_\nu\} =- \eta_{\mu\nu}$. The Hamiltonian $H_{\bar b}$ yields some of equations of motion that are {\it same}
as the ones that emerge from $H_b$. The following equations of motion are different, however, from (17), namely;  
\begin{eqnarray}
&&\dot {\Pi}_{(\beta)} = 2\, \bar b\, \bar\beta + 2\, \beta\, 
\bar\beta^2, \quad \dot {\Pi}_{(\bar\beta)} = 2\, \bar b\,\beta 
+ 2\, \beta^2\, \bar\beta, \quad \dot {\bar b} = \frac{1}{2} p^2 + 2 \gamma \chi,
\end{eqnarray}
which have been obtained by exploiting the general Heisenberg's equation of motion (16) with the Hamiltonian 
quoted in (13).

At this stage, it appears, as if, there is inconsistency amongst the equations of motion (10), (11), (17) and (19).
However, this is {\it not} the case. We may note that, with the inputs $\dot \beta = \dot {\bar\beta} = 0, 
\, \dot{\chi} = -i\, \gamma$, emerging from the Heisenberg's equation of motion and 
using the definition of the canonical momenta  in (14) and (15), one can prove
that the Euler-Lagrange equations of motion (10) and (11) reduce to the Heisenberg's equation 
of motion (17) and (19). Thus, we conclude that there is {\it no}
inconsistency between the equations of motion that emerge from the Lagrangian and Hamiltonian formulations.

\section{Time-evolution invariance of CF-type condition: Hamiltonian approach}

As is obvious from our discussions in the previous sections, the CF-type condition
$(b + \bar b + 2\,\beta\,\bar\beta = 0)$ emerges in {\it two steps}
from the coupled (but equivalent) Lagrangians (7) and (8) as well as from the Hamiltonians (12) and (13). 
However, this derivation is clumsy, as it is derived from, apparently {\it different} looking Lagrangians and Hamiltonians
which are equivalent only on a super world-line, defined by the CF-type restriction, in
the $D$-dimensional target spacetime manifold. To derive the CF condition in one stroke, we add the
Lagrangians in (7) and (8) to redefine the following new Lagrangian
\begin{eqnarray}
L^{(2)} = \frac{1}{2} \big( L^{(1)}_b + L^{(1)}_{\bar b} \big) = L_0 + L_g + L_{extra},
\end{eqnarray} 
where $L_0$ is given in equation (1) and the explicit expressions for $L_g$ and $L_{extra}$,
(in terms of the variables of the theory) are
\begin{eqnarray}
L_g &=& - i\, \dot {\bar c}\, \dot c + 2\,i\,(\beta\,\bar c - \beta\, c)\,\dot{\chi} 
+ 2\,(e\,\chi - \beta\, \bar c + \bar\beta\,c)\,\gamma + \beta^2\,\bar\beta^2 \nonumber\\ 
&+& (2\,i\, \bar c\, \chi - e \, \bar\beta) \, \dot {\beta} 
- (2\,i\, c\,\chi - e \, \beta)\, \dot{\bar\beta}, \nonumber\\
L_{extra} &=& \big(\frac{b - \bar b}{2}\big)\, \dot e + \big(\frac{b^2 + \bar b^2}{2} \big) + (b + \bar b)\, \beta\, \bar\beta.
\end{eqnarray}
We note, in passing, that the other linearly independent combination 
$L^{(3)} = \frac{1}{2} \big( L^{(1)}_b - L^{(1)}_{\bar b}\big)$
is not interesting for our discussions because ``$L_0$" cancels out along with the other 
useful as well as dynamically important terms.

Even at this stage, the Lagrangian (20) does not lead to the derivation of CF-type restriction in {\it one} step.
To corroborate the above statement, we observe that the following equations of motion emerge from (20), namely;
\begin{eqnarray}
b + \bar\beta\, \beta = -\frac{\dot e}{2}, \qquad\qquad  \bar b + \bar\beta\, \beta = \frac{\dot e}{2},
\end{eqnarray}
which, ultimately, lead to two linearly independent relationships between the Nakanishi-Lautrup 
type auxiliary variables `$b$' and `$\bar b$' as:
\begin{eqnarray}
b + \bar b = -2\,\beta\, \bar\beta,\qquad\qquad b - \bar b = -\dot e.
\end{eqnarray}
Thus, it is clear that we obtain CF-type condition $b + \bar b + 2\,\beta\,\bar\beta = 0$ in two steps from the Lagrangian (20), too.
To demonstrate the time-evolution invariance of CF-type condition, it is essential that we should obtain the above restriction, 
in one-stroke, from the appropriate Lagrangian and we should obtain the 
suitable Hamiltonian, corresponding to this Lagrangian, for our further theoretical analysis.

Towards this goal in mind, we redefine the following linearly independent variables
from the Nakanishi-Lautrup type auxiliary variables, namely;
\begin{eqnarray}
B = \frac{b + \bar b}{2}, \;\qquad \; \bar B = \frac{b - \bar b}{2}\; \Longrightarrow \;\frac {b^2 + {\bar b}^2}{2} = B^2 + {\bar B}^2.
\end{eqnarray}
This leads us to obtain the explicit expression for $L_{extra}$ as:
\begin{eqnarray}
L_{extra} = \bar B\, \dot e + \big(B^2 + \bar B^2 \big) +  2\,B\, \beta\, \bar\beta + e\,(\beta\,\dot{\bar\beta} - \bar\beta\, \dot{\beta}).
\end{eqnarray}
The Euler-Lagrange equation of motion, w.r.t. B, yields $B + \beta\,\bar\beta = 0$ 
which is nothing but the CF-type restriction ($ b + \bar b + 2\, \beta\,\bar\beta = 0$). 
The other equations of motion, emerging from the Lagrangian $L^{(2)}$, are as follows:
\begin{eqnarray}
&&\ddot c + 2\,\beta\, \dot{\chi} + 2\,i\, \beta\,\gamma + 2\, \chi\, \dot{\beta} = 0, \qquad 
\ddot {\bar c} + 2\,\bar\beta\, \dot{\chi} + 2\,i\, \bar\beta\,\gamma + 2\, \chi\, \dot{\bar\beta} = 0, \nonumber\\
&& 2\,i\big( \bar\beta\, \dot{c} - \beta\, \dot{\bar c}\big) = i\, p \cdot \psi + 2\, e\, \gamma,\quad
\dot {\bar B} + \frac{1}{2}\, p^2 + \big( \bar\beta \,\dot{\beta} - \beta\, \dot{\bar\beta}) - 2\, \chi\,\gamma = 0, \nonumber\\
&& \dot {\psi}^\mu = \chi\, p^\mu, \qquad \dot e \, \bar\beta + 2\, e\, \dot{\bar\beta} = 2\, \bar c\,\gamma + 2\,i\,\dot{\bar c}\,\chi
- 2\, \beta\, \bar\beta^2 - 2\, B\, \bar\beta, \nonumber\\
&& e \, \dot{\beta} = 2\,c\,\gamma + 2\,i\,\dot{c}\,\chi + 2\, \bar\beta\, \beta^2 
- 2\, B\,\beta,\qquad \bar B = -\frac{\dot e}{2},\nonumber\\ 
&&\dot {x}^\mu = e\,p^\mu - i\,\chi\,\psi^\mu, \qquad \dot {p}^\mu = 0, \qquad e\,\chi = \beta\, \bar c - \bar\beta\, c.
\end{eqnarray}
Thus, the appropriate Lagrangian for our present analysis, in terms of the auxiliary variables $B$ and $\bar B$, is
\begin{eqnarray}
L_B &=& p_\mu\, \dot x^\mu - \frac{e}{2}\, p^2 + \frac{i}{2}\, \psi_\mu\, \dot \psi^\mu 
+ i \,\chi\, p_\mu\, \psi^\mu + \bar B \,\dot e + (B^2 + {\bar B}^2) + 2\, B\,\beta \bar \beta \nonumber\\
&-& i \,\dot {\bar c}\,\dot c + 2\,i\,(\beta\, \bar c - \bar\beta\, c)\,\dot{\chi}
+ 2\, (e\,\chi - \beta\, \bar c + \bar\beta\, c)\,\gamma
+ \beta^2\, {\bar\beta}^2 \nonumber\\ &+& (2\,i\,\bar c\,\chi - e\,\bar\beta)\,\dot{\beta} 
- (2\,i\,c\,\chi - e\,\beta)\,\dot{\bar\beta},
\end{eqnarray}
which leads to the definition of non-vanishing canonical conjugate momenta as quoted in (14). The vanishing canonical
momenta are $\Pi_{(B)} \approx 0$ and $\Pi_{(\gamma)} \approx 0$ which are the primary constraints{\footnote{ We 
discuss about these issues in our fifth section  (in somewhat more detail). }}
on the theory and both of them are canonically conjugate to the auxiliary variables $B$ and $\gamma$, respectively.

The canonical Hamiltonian ($H_B$), that is derived for the appropriate Lagrangian ($L_B$) using Legendre
transformation, is as follows
\begin{eqnarray}
H_B &=&  i\, \Pi_{(\bar c)}\,\Pi_{(c)} + \frac{e}{2}\, p^2 - i \,\chi\, p_\mu\, \psi^\mu 
- \beta^2\, {\bar\beta}^2 - 2\, B\,\beta \bar \beta - (B^2 + {\bar B}^2)  \nonumber\\
&-& 2\, e\,\chi \,\gamma + i\, \Pi_{(\chi)}\,\gamma,
\end{eqnarray}
where the canonical conjugate momenta, corresponding to the variables in (27), are same as (14) except the following:
\begin{eqnarray}
\Pi_{(e)} = \bar B,\qquad \Pi_{(\beta)} = 2\,i\, \bar c\,\chi - e\,\bar\beta, 
\qquad \Pi_{(\bar\beta)} = - 2\,i\, c\,\chi + e\,\beta.
\end{eqnarray}
It is now straightforward to check that
\begin{eqnarray}
\dot{\Pi}_{(B)} = -i\,[ \Pi_{(B)}, \, H_B] = 2\,(B + \beta\,\bar\beta) = 0,\qquad [B + \beta\,\bar\beta, H_B] = 0.
\end{eqnarray}
The above equations establish the derivation of the CF condition from $\dot{\Pi}_{(B)} \approx 0$ as the
secondary constraint (i.e. $B + \beta\, \bar\beta = 0$) of the theory. Furthermore, we also obtain the
time-evolution invariance of the CF-type condition as the above secondary constraint commutes with the Hamiltonian of the theory.

\section{Conclusions}
In our present investigation, we have derived the appropriate Lagrangian that yields the CF-type restriction as an 
Euler-Lagrange equation of motion by redefining the auxiliary variables (e.g. $B$ and $\bar B$) in terms of the
Nakanishi-Lautrup auxiliary variables (e.g. $b$ and $\bar b$). As it turns out, this Lagrangian is endowed with
{\it three} primary constraints $\Pi_{(\gamma)} \approx 0,\, \Pi_{(B)} \approx 0$ and $\Pi_{(\bar B)} \approx 0$
as there are no ``velocity" terms corresponding to the variables $\gamma,\, B$ and $\bar B$.
It is essential, however, to lay emphasis on the fact that variable $\bar B$ is somewhat special because it is canonical
conjugate momentum for the variable $e$ [cf. eqs.(29)]. Thus, effectively, there are only {\it two} 
primary constraints.

In the Lagrangian formulation, the time-evolution invariance of the CF-type restriction does not appear in a 
straightforward manner even though we use the CF-type restriction in proving the absolute anticommutativity of the (anti-)BRST
symmetries of the theory. We have, in our present investigation, derived the appropriate Hamiltonian for the theory and demonstrated
that the CF-type condition is nothing other than the secondary constraint on the theory that 
emerges when we demand the time-evolution invariance
($\dot {\Pi}_{(B)} \approx 0$) of the primary constraint $\Pi_{(B)} \approx 0$. This secondary constraint, 
as it turns out,
commutes with the Hamiltonian of the theory thereby proving its time-evolution invariance.
This justifies the imposition of CF-type condition on the theory as it remains the {\it same} during the time-evolution.

In the above context, it is important to point out that the time-evolution invariance of $\Pi_{(\gamma)}$
($\dot{\Pi}_{(\gamma)} \approx 0$) produces $e\,\chi = \beta\,\bar c - \bar\beta\, c$ which is one of the equations 
of motion that emerges out from the Lagrangian (27). However, in the language of constraint
analysis, the equation $e\,\chi = \beta\,\bar c - \bar\beta\, c$ is {\it also} a secondary constraint. We do {\it not}
perform here a detailed constraint analysis.

We have established the time-evolution invariance of the CF-type restriction in the context of 4D Abelian 2-form gauge theory
[9] and achieved the same goal for our present SUSY toy model of the 1D free spinning relativistic particle.
It would be nice endeavor to carry out these kinds of exercise in the context of other physically interesting models so that
we could lay our method of analysis on the firmer footings. We are intensively involved in this endeavor at the moment
and our result would be reported later [10].

%\acknowledgments
%One of us (A.S.) is grateful to the UGC, Govt. of India, New Delhi, for financial
%support under RFSMS scheme, under which, this investigation has been performed.

\end{document}